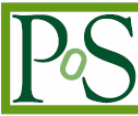

# Search for Leptoquarks and Compositeness at D0

**Maxim Titov**[*]

*Albert-Ludwigs University of Freiburg, Physics Institute,*
*Hermann-Herder Str.3, 79104, Freiburg, Germany*
*E-mail:* `maxim.titov@physik.uni-freiburg.de`

**(on behalf of the D0 Collaboration)**

In this paper searches are presented for the pair production of first and second generation scalar leptoquarks and limits are given on the quark-lepton compositeness scale $\Lambda$ from proton-antiproton collision data at a center-of-mass energy 1.96 TeV, collected with Run II D0 Detector in 2002-2004.

No evidence for a leptoquark signal has been observed in the topologies arising from $LQ_1LQ_1 \rightarrow ejej$ and $LQ_1LQ_1 \rightarrow ej\nu j$, using an integrated luminosity of 252 pb$^{-1}$, and in the $LQ_2LQ_2 \rightarrow \mu j\mu j$ channel, using a data sample corresponding to 294 pb$^{-1}$. From the upper bounds on the product of cross section times branching ratio $\beta = Br(LQ \rightarrow lj)$, a lower mass limits of $M_{LQ1} > 241$ GeV and $M_{LQ2} > 247$ GeV for the first and second LQ generation are set for $\beta=1$. These results, combined with those obtained by D0 in Run I at a center-of-mass energy of 1.8 TeV, allow to exclude scalar LQ masses up to 256 GeV and 251 GeV (for $\beta=1$) for the first and second generation, respectively. These limits are the most stringent up-to-date constraints on the first and second generation LQ masses, which are independent of the strength of the lepton-quark-leptoquark ($\lambda_{LQ}$) Yukawa coupling.

The dilepton mass spectra in $pp \rightarrow l^+l^- + X$ interactions are studied using dielectron (dimuon) data samples, corresponding to an integrated luminosity of 271 pb$^{-1}$ (406 pb$^{-1}$). The mass spectra being a probe for new physics are examined for new interactions of quarks and leptons from a common composite structure. No excess of events is found over the expectation from Standard Model processes. The current experimental lower limits on the compositeness scale, $\Lambda$, vary, for different chirality channels, from 3.6 to 9.1 TeV for the (eeqq) and from 4.2 to 9.8 TeV for the ($\mu\mu qq$) contact interaction. The D0 Run II limits on ($\mu\mu qq$) are currently the most stringent bounds for the quark-muon compositeness scale.

[*] Speaker

# 1. Scalar Leptoquark Phenomenology

Leptoquarks (LQ) are color-triplet bosons, carrying both baryon and lepton quantum numbers and fractional electric charge. They appear in many extensions beyond the standard model (SM) such as composite models with quark and lepton substructure, extended technicolor, SUSY with R-parity violations and grand unified theories (GUT), based on gauge groups SU(5), SO(10) with Pati-Salam SU(4) color symmetry, SU(15) and superstring-inspired E6 models [1]. The general classification of leptoquark states, proposed by Buchmuller, Ruckl and Wyler, contains 7 scalar and 7 vector leptoquarks [2]. At the Tevatron, scalar leptoquarks would be pair produced dominantly through quark-antiquark annihilation (for $M_{LQ}$>100 GeV) and gluon fusion [3]. Since the gluon-leptoquark interactions are determined by the non-Abelian $SU(3)_C$ gauge symmetry of scalar QCD, the cross section for scalar LQ pair production is essentially independent of the lepton-quark-LQ Yukawa coupling ($\lambda_{LQ}$), which contributes only ~ 1% (for $\lambda_{LQ} \sim \lambda_{em}$) of the total cross section in t-channel lepton exchange of qq-annihillation. The cross section coincides with those of squark-pair production in the limit of large gluino masses [3]. While the scalar LQ pair production at the Tevatron does not depend on its electroweak properties, leptoquarks are generally assumed to have a family diagonal coupling (LQ couples to a single lepton/quark generation only – to avoid flavor-changing neutral currents (FCNC) or lepton flavor violation) and chiral coupling (to obey bounds arising from atomic parity violation and to avoid anomalously large contributions to $\pi \rightarrow e\nu$ decays). Finally, LQ couplings to fermions are assumed to be baryon and lepton number conserving, to avoid rapid proton decays.

# 2. First and Second Generation Scalar Leptoquarks Limits

The experimental signatures for the LQ pair production at the Tevatron are two jets accompanied by either two leptons (ljlj), lepton and missing transverse energy $E_T^{miss}$ (lνjj), or $E_T^{miss}$ (ννjj), with each pair of jet +lepton invariant masses being equal to the leptoquark mass.

D0 has searched for the first generation scalar leptoquarks in the $LQ_1LQ_1 \rightarrow$ ejej and ejνj channels based on Run II integrated luminosity of 252 pb$^{-1}$. The (ejej) analysis requires two electrons with transverse energy $E_T^e$>25 GeV and at least two jets with $E_T^j$>20 GeV. The (eνjj) data sample is selected with one electron ($E_T^e$>35 GeV), more than two jets with $E_T^j$>25 GeV, and $E_T^{miss}$>30 GeV. The major background that mimic (ejej) decay of LQ pair are Z/γ*+2 jets, tt production and misidentified multi-jet events, while the primary backgrounds to the (eνjj) final state are W+2 jets, multijet and tt production. A significant discrimination between the LQ signal and the major Z/γ*+jets and W+jets backgrounds is achieved by a cut on the scalar transverse energy sum: $S_T = E_T^{j1} + E_T^{j2} + E_T^{e1} + E_T^{e2}$ > 450 GeV for the (ejej) and $S_T = E_T^{j1} + E_T^{j2} + E_T^{e1} + E_T^{miss}$ > 330 GeV for the (eνej) final state. As no excess of data above background is found after the final $S_T$ cut, a 95% C.L. limit on the first-generation LQ mass is set at $M_{LQ1}$> 241 GeV ($M_{LQ1}$> 218 GeV) for β=Br(LQ$\rightarrow$lj)=1 (β=1/2), based on Run II combination of (ejej) and (eνjj) channels, using a Bayesian likelihood technique [4]. The D0 Run II and combined Run I+II lower limits, on the first-generation leptoquark mass as a function of β, are shown in Fig.1 [5].

<pre>                    **arXiv: hep-ex/0512006 v1 3 December 2005**</pre>

D0 has also performed a similar search for second generation scalar LQ2 decaying through LQ2LQ2→μjμj. The Run II integrated luminosity of the data used in this analysis is 294 pb$^{-1}$. After the initial event selection, which requires the presence of two muons with $p_T$>15 GeV and two jets with $E^j_T$>25 GeV, the background is dominated by the Z/γ*+jets and tt production. The LQ2LQ2→μjμj events are expected to have both high di-muon masses ($M_{\mu\mu}$) and large values of $S_T=E_T^{j1}+E_T^{j2}+p_T^{\mu1}+p_T^{\mu2}$. Therefore, the sensitivity to LQ decays is studied using the two-dimensional distribution ($S_T$ versus $M_{\mu\mu}$) divided into four signal bins, in order of increasing signal over background ratio (S/B). No excess of data over SM background was found. The D0 Run II lower limit at the 95% C.L., calculated by treating the four signal bins as individual channels and combining them using the modified frequentist approach [6], on the mass of the second generation scalar leptoquark is $M_{LQ2}$>247 GeV ($M_{LQ2}$>182 GeV) for β=1 (β=1/2), respectively. The combined D0 Run I+II lower limit for a combination of channels (μjμj) and (νjμj) (νjμj was analyzed only at Run I) for the second-generation scalar LQ is $M_{LQ2}$>251 GeV ($M_{LQ2}$>204 GeV) for β=1(β=1/2). The excluded parameter regions, which are independent from $\lambda_{LQ,}$ are shown in Fig.1 (right).

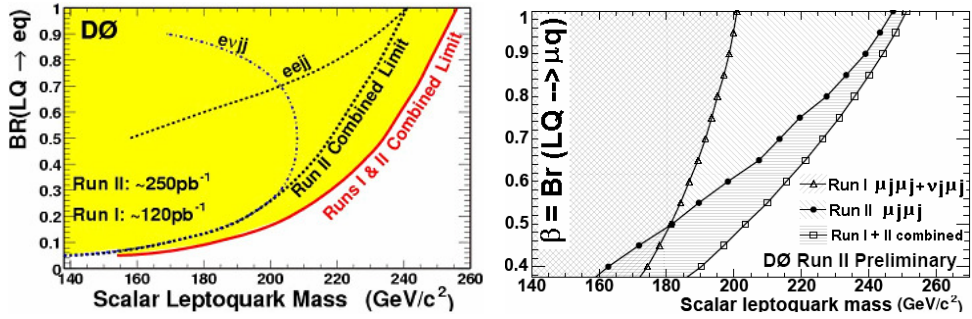

Fig. 1 The D0 Run II and combined Run I+II excluded regions (filled/shaded area) in the β=Br (LQ →lj) versus LQ mass plane for the first-generation (left) and second-generation (right) scalar leptoquarks.

## 3. Contact Interactions and Compositeness Limits

A general framework to assess new physics, appearing from the interference of any new particle field, associated to a characteristic energy scale ($\Lambda^2$>>**s**), with the γ and Z fields of the SM, is a concept of four-fermion point-like (ff)(ff) contact interactions (CI) [7]. The most general Lagrangian for CI between leptons and quarks can be written in the form:

$$L = \Sigma_q \Sigma_{A,B} \ \eta_{AB}^q (l_A\gamma_\mu l_A)(q_B\gamma_\mu q_B); \quad \eta_{AB}^q = \varepsilon_{AB}(g/\Lambda_{AB}^q)^2$$

where $\eta_{AB}$ describes the chiral structure of the interaction (A,B=L,R), and the $\Sigma_q$ extends over all up-type and down-type quarks and antiquarks, g/Λ is an effective coupling over the scale Λ, $\varepsilon_{AB}$=-1(+1) defines constructive (destructive) interference of the CI with the SM. Any new phenomena, such as compositeness (Λ ~ scale of composite objects), or the exchange of heavy leptoquarks (g/Λ~$\lambda_{LQ}$/$M_{LQ}$), can be described by an appropriate choice of $\eta_{AB.}$ Compositeness models postulate common constituents of the SM fermions and a new strong dynamics (`metacolor') that bind these constituents. Experimentally, quark compositeness is the most sensitive to the deviation in production of high-transverse momentum jets relative to SM

<pre>                                                                         PoS(HEP2005)314</pre>

predictions. Quark-lepton compositeness would modify the SM Drell-Yan cross section for lepton pair production in the high mass region above the Z peak.

The $Z/\gamma^* \to ee$ ($Z/\gamma^* \to \mu\mu$) invariant mass spectra from 271 pb$^{-1}$ (406 pb$^{-1}$) of collisions, taken by the D0 detector in Run II, are used to measure the shape of the cross sections and to search for continuum anomalies with respect to the SM predictions. To avoid biases from backgrounds, the data selection is optimized for purity, by requiring two electrons of $p_T^e > 25$ GeV in the dielectron, and two isolated muons with $p_T^\mu > 15$ GeV and $M_{\mu\mu} > 50$ GeV in the dimuon analysis. The major backgrounds for $Z/\gamma^* \to ee$ - QCD multijet and $\gamma$+jet events, where both objects fake electrons, are determined directly from the data. No evidence for physics beyond the SM is observed both in $e^+e^-$ and $\mu^+\mu^-$ data. The lower limits on the quark-lepton compositeness scale are set independently for each chirality channel of the CI Lagrangian, using a Bayesian analysis of the shape of the $M_{ee}$ mass distributions of events or from the two-dimensional distributions ($M_{\mu\mu}$ vs. cos $\theta^*$). The D0 Run II lower bounds on the quark-electron and quark-muon compositeness scale $\Lambda$ range from 3.6 to 9.1 TeV for (eeqq) and from 4.2 to 9.8 TeV for ($\mu\mu$qq) CI (see Fig. 2). The current summary of 95% C.L. lower limits on the CI energy scale $\Lambda$ from LEP, HERA and Tevatron collider experiments is also shown in Fig.2.

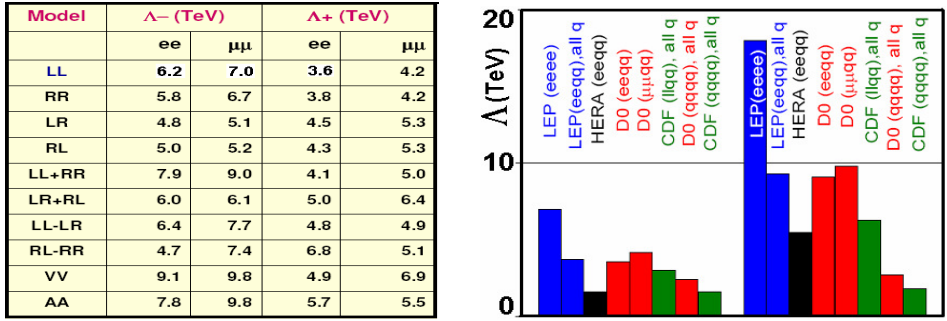

| Model | $\Lambda^-$ (TeV) | | $\Lambda^+$ (TeV) | |
|-------|-----|-----|-----|-----|
|       | ee  | $\mu\mu$ | ee | $\mu\mu$ |
| LL    | 6.2 | 7.0 | 3.6 | 4.2 |
| RR    | 5.8 | 6.7 | 3.8 | 4.2 |
| LR    | 4.8 | 5.1 | 4.5 | 5.3 |
| RL    | 5.0 | 5.2 | 4.3 | 5.3 |
| LL+RR | 7.9 | 9.0 | 4.1 | 5.0 |
| LR+RL | 6.0 | 6.1 | 5.0 | 6.4 |
| LL-LR | 6.4 | 7.7 | 4.8 | 4.9 |
| RL-RR | 4.7 | 7.4 | 6.8 | 5.1 |
| VV    | 9.1 | 9.8 | 4.9 | 6.9 |
| AA    | 7.8 | 9.8 | 5.7 | 5.5 |

Fig. 2. (Left) The 95% C.L. D0 Run II lower limits on the quark-electron and quark-muon compositeness scale for constructive ($\Lambda^-$) and destructive ($\Lambda^+$) CI interference with the SM Lagrangian. (Right) Summary of the 95% C.L. lower limits on the electron (eeee), quark (qqqq) and quark-lepton (llqq) compositeness scale $\Lambda$ of contact interactions from LEP [8], HERA [9] and Fermilab (CDF[10], D0) Experiments. The lower and upper values of $\Lambda$ range, among the different chirality channels, are shown. (For LEP, (eeqq) CI are assumed to couple to all quark flavors with equal strength. The CDF Run I (llqq) limits assume that leptons couple symmetrically to u-type (u,c,t) and d-type (d,s,b) quarks. All quarks families are assumed to be composite for (qqqq) CI.